\documentclass[conference, twocolumn]{IEEEtran}

\usepackage[utf8]{inputenc} 	% UTF-8

\usepackage{amssymb}
\usepackage{bbm} % For indicator function
\usepackage{graphicx}
\usepackage{color,psfrag}

\usepackage{MnSymbol}
\usepackage[normalem]{ulem}

\usepackage{tikz}
\usepackage{siunitx}
\usepackage{cite} % Combining multiple citations
\newcommand{\e}[2]{{\mathbb E}_{#1}\left[ #2 \right]}
\newcommand{\s}[2]{{\frac{1}{{#1}}\sum_n^{#1}} {#2}}
\newcommand{\q}[2]{{\mathcal Q}_{#1}\left( #2 \right)}
\newcommand{\p}{\mathcal P}
\newcommand{\sub}[1]{_{\text{#1}}}

\newcommand{\pc}{\text{P}\sub{c}}
\newcommand{\pcd}{\bar{\text{P}}\sub{c}}

\newcommand{\preg}{P\sub{cont}}
\newcommand{\xreg}{x\sub{cont}}
\newcommand{\prcvd}{P\sub{rcvd}}
\newcommand{\yrcvd}{y\sub{rcvd}}
\newcommand{\ptran}{P\sub{tran}}
\newcommand{\xtran}{x\sub{tran}}
\newcommand{\pp}{P\sub{p}}
\newcommand{\ps}{P\sub{s}}
\newcommand{\yp}{y\sub{p}}
\newcommand{\ys}{y\sub{s}}
\newcommand{\nap}{w\sub{p}}
\newcommand{\nas}{w\sub{s}}
\newcommand{\ite}{\theta\sub{I}}
\newcommand{\rs}{R\sub{s}}
\newcommand{\ers}{\e{}{\rs}}
\newcommand{\gp}{g\sub{p}}
\newcommand{\gs}{g\sub{s}}
\newcommand{\ap}{\alpha\sub{p}}
\newcommand{\as}{\alpha\sub{s}}
\newcommand{\npp}{\sigma^2\sub{p}}
\newcommand{\nps}{\sigma^2\sub{s}}
\newcommand{\npu}{\Delta\sigma^2}
\newcommand{\fsam}{f\sub{s}}

\newcommand{\fpp}{F_{\pp}}

\newcommand{\dK}{f_{\gp \ap/\prcvd}}
\newcommand{\dpp}{f_{\pp}}
\newcommand{\dpreg}{f_{\preg}}
\newcommand{\drs}{f_{\rs}}

\DeclareMathOperator*{\maxi}{max}

\usepackage[final=true]{hyperref}
\hypersetup{
	pdfauthor = {Ankit Kaushik et al.},
	pdftitle = {ICC 2015},
	pdfsubject = {ICC 2015},
	pdfcreator = {PDFLaTeX with hyperref package},
	pdfproducer = {PDFLaTeX}
}

\allowdisplaybreaks

\usepackage{flushend}

\begin{document}
\title{Estimation-Throughput Tradeoff for Underlay Cognitive Radio Systems}
\author{\IEEEauthorblockN{Ankit Kaushik\IEEEauthorrefmark{1}, Shree Krishna Sharma\IEEEauthorrefmark{2}, Symeon Chatzinotas\IEEEauthorrefmark{2},  Bj\"orn Ottersten\IEEEauthorrefmark{2}, Friedrich Jondral\IEEEauthorrefmark{1}}
\IEEEauthorblockA{\IEEEauthorrefmark{1}Communications Engineering Lab, Karlsruhe Institute of Technology (KIT), Germany}
\{ankit.kaushik, friedrich.jondral\}@kit.edu
\IEEEauthorblockA{\IEEEauthorrefmark{2}SnT - securityandtrust.lu, University of Luxembourg, Luxembourg}
\{shree.sharma, symeon.chatzinotas, bjorn.ottersten\}@uni.lu
}

% make the title area
\maketitle
\thispagestyle{empty}
\pagestyle{empty}

\begin{abstract}
Understanding the performance of cognitive radio systems is of great interest. To perform dynamic spectrum access, different paradigms are conceptualized in the literature. Of these, Underlay System (US) has caught much attention in the recent past. According to US, a power control mechanism is employed at the Secondary Transmitter (ST) to constrain the interference at the Primary Receiver (PR) below a certain threshold. However, it requires the knowledge of channel towards PR at the ST. This knowledge can be obtained by estimating the received power, assuming a beacon or a pilot channel transmission by the PR. This estimation is never perfect, hence the induced error may distort the true performance of the US. Motivated by this fact, we propose a novel model that captures the effect of channel estimation errors on the performance of the system. More specifically, we characterize the performance of the US in terms of the estimation-throughput tradeoff.
Furthermore, we determine the maximum achievable throughput for the secondary link. Based on numerical analysis, it is shown that the conventional model overestimates the performance of the US.
%\imp{important statement} \\
%\ur{urgent or critical} \\
%\ns{not sure if that is true} \\
%\ws{wrong statement} \\
%\fl{flow of the paper} \\
%\un{unclear statement or argument}
\end{abstract}
%%%%%%%%%%%%%%%%%%%%%%%%%%%%%%%%%%%%%%%%%%%%%%%%%%%%%%%%%%%%%%%%%%%%%%%%%%%%%%%%%%%%%%%%%

\section{Introduction}%%%%%%%%%%%%%%%%%%%%%%%%%%%%%%%%%%%%%%%%%%%%%%%%%%%%%%%%%%%%%%%%%%%%%%%%%%%%%%%%%%%%%%%%%
%\subsection{Background}
Cognitive radio communication is considered as a potential solution in order to address the spectrum scarcity problem of future wireless networks. The available cognitive radio paradigms in the 
literature can be categorized into interweave, underlay and overlay \cite{Goldsmith09}. In Interweave Systems (IS), the Secondary Users (SUs) utilize the primary licensed spectrum opportunistically by exploiting spectral holes in different domains such as time, frequency, space, polarization, etc, whereas in Underlay Systems (US), SUs are allowed to use the primary spectrum as long as they respect the interference constraints of the Primary Receivers (PRs). On the other hand, Overlay Systems (OS) allow the spectral coexistence of two or more wireless networks by employing advanced transmission and coding strategies. Of these mentioned paradigms, this paper focuses on the performance analysis of the USs in terms of the estimation-throughput tradeoff considering transmit power control at the Secondary Transmitter (ST) with the help of the employed channel estimation technique.
%Goldsmith \textit{et .al} \cite{Goldsmith09}

\subsection{Motivation}
The main advantage of the US over the IS comes from the fact that the former system allows the SUs to transmit in a particular frequency band even if a Primary User (PU) is operating in that band. Further, PRs have some interference tolerance capability which is totally neglected in IS systems \cite{Jiang13, Sharma14}. It should be noted that the interference caused by the STs is not harmful to the PRs all the time since it becomes harmful only if the interference exceeds the Interference Threshold (IT) of the PR. In this context, USs make better utilization of the available frequency resources in spectrum sharing scenarios. This is the main motivation behind studying underlay scenario in this paper.

In the literature, the performance analysis of the US is limited to knowledge of the channel. With its knowledge, ST operates at a transmit power such that the IT is satisfied at the PR. 
Although several existing contributions have considered power control in the USs, the channel estimation aspect has received less attention and the performance analysis of power control-based USs considering estimation errors is still an open problem. 
In a realistic scenario, 
%this knowledge is acquired by CR through estimation. The estimation in presence of noise leads to a certain variation in the performance. If not considered in the model, these variations may distort the performance parameters. Thereby it becomes difficult to capture the true performance  of the system. 
direct estimation of the channel is not possible, rather this has to be done indirectly by listening to a beacon or a neighbouring pilot channel transmitted by the PR. That is, the ST employs a power based estimation technique by evaluating the received power. Furthermore, implementing power control based on the estimated received power requires knowledge of its noise power, which is rarely perfect \cite{Tan08}. Hence, in order to characterize the true performance of the US, it is important to include these aspects in the model.  

Further, sensing-throughput tradeoff has been considered as an important performance metric while analyzing the performance of the ISs \cite{Liang08, Juarez11, Sharkasi12}. However, for the USs, the situation is different since the ST is involved in estimating the received power instead of simply detecting the presence or absence of the primary signals. With the inclusion of estimation error, US intends to operate at a suitable estimation time such that the probability of confidence remains above the desired level. In this sense, similar to the sensing-throughput tradeoff in ISs, it is evident that there exists a tradeoff between estimation time and secondary throughput in USs. In this context, this paper studies the estimation-throughput tradeoff in USs considering estimation errors while estimating the channel to the PR. 

The performance of IS via sensing-throughput tradeoff has been characterized by Liang \textit{et. al.} in \cite{Liang08}. According to \cite{Liang08}, the objective is to maximize throughput at the ST subject to the sharing constraint set by regulation for primary system. 
%The sharing constraint required the probability of detection to sustain a desired level. 
We intend to derive a similar expression for the ST acting as an US. %Analogous to IS, US protect the primary system by regulating their transmit power below a certain IT at PR. To pursue this, CR has to estimate the state of the channel. The estimation of the channel state is performed by listening to received power over a neighbouring pilot channel or the beacon sent by the PR. As a consequence to estimation, an error is induced into the system. 
The objective of the ST as an US is to maximize the throughput such that a desired probability of confidence is sustained for a certain accuracy. This accuracy can be defined using a set of confidence intervals.

\subsection{Contributions}
%\begin{itemize}
%\item 
We propose a realistic model, according to which an ST estimates the channel. As a consequence to that, we investigate the true performance of the US. 
%\item 
In order to perform channel estimation, we employ a power based estimation technique. Additionally, we include the noise uncertainty in the model to obtain the best and worst performance bounds for the US. 
%\item 
To analyze the performance for the proposed model, we characterize the expected throughput attained at the Secondary Receiver (SR) and the probability of confidence at the PR. Consequently, we examine the estimation-throughput tradeoff for the US. As a result of the analysis, we propose a power control scheme at the ST subject to probability of confidence constraint at the PR. 
%\item 
Finally, we examine the performance of the proposed model with path loss and fading channels. Most importantly, the performance of the system is characterized based on exact analytical expressions.
%\item Characterizing the performance of US via estimation-throughput tradeoff. We consider the throughput $R(\tau)$ as function of estimation time $\tau$. $\pc(\mu, \tau)$ is characterized by the distribution function. 
%\item In practice, the estimation of the channel state from the received power requires the knowledge of noise power. We modify the above expression to incorporate the noise uncertainty into the estimation-throughput tradeoff.
%\item Finally, we compare the two performance of the IS and US systems. (Nice to have!)
%\item An expression of expected throughput $\e{h}{R}$ for a fixed estimation time and accuracy of the system. (Nice to have!)
%\end{itemize}
%\subsection{Organization}
%\imp{Can be done at the end.}

The rest of the paper is organized as follows: Section \ref{sec:sys_mod} describes the system model that includes the underlay scenario and the signal model. Section \ref{sec:et_ana} discusses the estimation-throughput analysis for the US and provides analytical expressions to characterize the performance parameters for the system. Section \ref{sec:num_ana} analyzes the numerical results based on the expressions obtained for the path loss and fading channels. Finally, Section \ref{sec:conc} concludes the paper. 
%%%%%%%%%%%%%%%%%%%%%%%%%%%%%%%%%%%%%%%%%%%%%%%%%%%%%%%%%%%%%%%%%%%%%%%%%%%%%%%%%%%%%%%%%
\section{System Model} \label{sec:sys_mod}
%%%%%%%%%%%%%%%%%%%%%%%%%%%%%%%%%%%%%%%%%%%%%%%%%%%%%%%%%%%%%%%%%%%%%%%%%%%%%%%%%%%%%%%%%

\subsection{Underlay scenario}
\begin{figure}[!t]
\centering
\includegraphics[width = 0.85 \columnwidth]{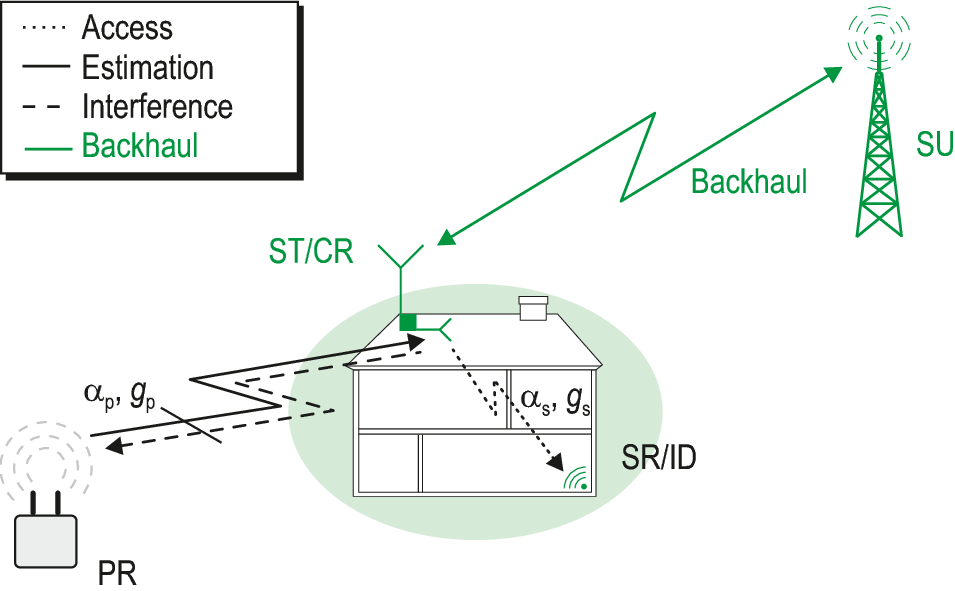}
\caption{A scenario demonstrating the underlay paradigm.}
\label{fig:scenario}
\vspace{-5mm}
\end{figure}

Cognitive Relay (CR) \cite{Kaushik14} characterizes a small cell deployment that fulfills the spectral requirements for Indoor Devices (IDs). \figurename~\ref{fig:scenario} illustrates a snapshot of a CR scenario to depict the interaction between the CR with PR and ID, where CR and ID represents the ST and SR respectively. In \cite{Kaushik14}, the challenges involved while deploying the CR as US were presented. However for simplification, a constant transmit power was considered at the CR. Now, we extend the analysis to employ power control at the ST. 
%\subsubsection{Medium access}

The medium access for the US is slotted, where the time axis is segmented into frames of length $T$. The frame structure is analog to periodic sensing in IS \cite{Liang08}. Unlike IS, US uses $\tau$ to estimate the received power, where $\tau (< T)$ corresponds to a time interval. To incorporate fading in the model, we assume that the channel remains constant for $T$. Hence characterized by the fading process, each frame witnesses a different received power. Therefore to sustain a desired probability of confidence, it is important to exercise estimation followed by power control for each frame. Thus, the time $T - \tau$ is utilized for data transmission with controlled power. 

To execute power control, ST has to consider the power received at the PR. This is done by listening to a beacon sent by the PR in the same band \cite{Jing10}. With the knowledge of power transmitted by the PR and using channel reciprocity \cite{Tse05}, ST is able to determine power received at the PR thereby controls the transmit power while sustaining IT. In case the primary system doesn't support the beacon transmission, the ST can listen to a neighbouring pilot channel for determining the received power. In order to apply channel reciprocity for the pilot channel based technique, it is assumed that center frequency separation between pilot channel and band of interest is smaller than the coherence bandwidth. %Hence, channel reciprocity can be applied also to the pilot channel based technique. In case the ST employs pilot channel based technique, 
For the mentioned techniques, we consider a proper alignment of the ST to the PR transmissions. Hence, during data transmission at the ST, the SR procures no interference from the PR, cf. \figurename~\ref{fig:scenario}. 

\subsection{Signal and channel model}
The received signal at the ST, transmitted by the PR cf. \figurename~\ref{fig:scenario}, is sampled with a sampling frequency of $\fsam$ and is given by
\begin{equation}
\yrcvd[n] = \sqrt{\gp \cdot \ap} \cdot \xtran[n] + \nas[n],
\label{eq:sys_mod_st}
\end{equation}
where $\xtran$ corresponds to a fixed discrete samples transmitted by the PR, $\gp \cdot \ap$ represents the power gain for the channel PR-ST and $\nas[n]$ is circularly symmetric complex Additive White Gaussian Noise (AWGN) at the ST. 
The transmitted power at PR is $\ptran = \s{\tau \fsam}{|\xtran[n]|^2}$, considering that $\tau \fsam$ $(= N)$ is the number of samples used for estimation. $\nas[n]$ is an independent identically distributed (i.i.d.) Gaussian random process with zero mean and variance $\e{}{|\nas[n]|^2} = \nps$. The power received at the ST is given as 
\begin{align}
\prcvd = \s{\tau \fsam}{ |\sqrt{\gp \ap} \xtran[n] + \nas|^2}.
\label{eq:prcvd} 
\end{align}
Analog to (\ref{eq:sys_mod_st}), the received signal at the PR, transmitted by the ST, is given by
\begin{equation}
\yp[n] = \sqrt{\gp \cdot \ap} \cdot \xreg[n] + \nap[n],
\label{eq:sys_mod_pr}
\end{equation}
and on the other side, the received signal at the SR follows 
\begin{equation}
\ys[n] = \sqrt{\gs \cdot \as} \cdot \xreg[n] + \nas[n],
\label{eq:sys_mod_sr}
\end{equation}
where $\xreg[n]$ is an i.i.d random process. $\preg = \s{(T - \tau) \fsam}{|\xreg[n]|^2}$ corresponds to controlled power at the ST. Further, $\gp \cdot \ap$ and $\gs \cdot \as$ represent the power gains for channel ST-PR and ST-SR, cf. \figurename~\ref{fig:scenario}. 
Considering (\ref{eq:sys_mod_pr}) and \ref{eq:sys_mod_sr}, the powers received at the PR and SR are evaluated as $\pp = \s{(T - \tau) \fsam}{|\yp[n]^2|}$ and $\ps = \s{(T - \tau) \fsam}{|\ys[n]^2|}$. Likewise (\ref{eq:sys_mod_st}), $\nap[n]$ and $\nas[n]$ represents circularly symmetric AWGN at PR and ST with zero mean and variance $\e{}{|\nap[n]|^2} = \npp$ and $\e{}{|\nas[n]|^2} = \nps$, correspondingly. %Consider that $\ptran$, $\preg$ and $\pp$ correspond to power for a given frame. 
%For (\ref{eq:sys_mod_sr}), we assume that the interference at SR from the PR via beacon or pilot channel is $< \nps$, which is a valid assumption considering the indoor scenario depicted in \figurename~\ref{fig:scenario}. 
%For (\ref{eq:sys_mod_sr}), we consider a perfect allignment of the ST to the beacon or pilot channe
%\subsubsection{Channel}

We consider that all transmitted signals are subjected to distance dependent path loss $\ap, \as$. The small scale fading gains $\gp, \gs$ are modelled as frequency-flat fading, thus, follow a unit-mean exponential distribution \cite{Tse05}. In the analysis, we consider the coherence time of the channels $\approx T$. But, there will be scenarios where the coherence time exceeds $T$, in such cases our characterization depicts a lower performance bound.

%%%%%%%%%%%%%%%%%%%%%%%%%%%%%%%%%%%%%%%%%%%%%%%%%%%%%%%%%%%%%%%%%%%%%%%%%%%%%%%%%%%%%%%%%
\section{Estimation-throughput analysis} \label{sec:et_ana}
%%%%%%%%%%%%%%%%%%%%%%%%%%%%%%%%%%%%%%%%%%%%%%%%%%%%%%%%%%%%%%%%%%%%%%%%%%%%%%%%%%%%%%%%%

\subsection{Conventional model}
According to the conventional model, ST as US is required to control its transmit power $\preg$ such that the received power $\pp$ at the PR is below IT ($\ite$) \cite{Xing07} given by
\begin{equation}
\pp = \gp \ap \preg \le \ite 
\label{eq:IT}
\end{equation}
%where $\alpha$ denotes the distance dependent path loss. $g\sub{p}$ represents the small-scale channel fading. 

With controlled power at the ST, the expected data rate at the SR is defined as
\begin{equation}
\e{\gs, \preg}{\rs} = \e{\gs, \preg}{\log_2 \left(1 + \frac{\gs \as \preg}{\nps} \right)}. 
\label{eq:Thr_cm}
\end{equation}
where $\e{\gs, \preg}{\cdot}$ represents the expectation over $\preg$ at ST and the channel gain $\gs$. 
\subsection{Proposed Model} 

To determine $\preg$ according to (\ref{eq:IT}), the conventional model considers that $\gp \ap$ is perfectly known at the ST, which is rarely the case. Hence, channel estimation must be included in the model. This however, opens up the following issues. First, the estimation of the channel cannot be carried out directly \cite{Tian12, Sharma13}. Second, the model must include the effect of estimation errors and noise uncertainty on the system's performance. In view of this, we consider these aspects in the proposed model. 
\subsubsection*{Power control}
We first obtain the information concerning the $\gp \ap$ by estimating the $\prcvd$. As a result, we determine $\preg$ directly from the $\prcvd$ based on the following expression 
\begin{align}
\preg = \frac{\ite K}{\prcvd}, \label{eq:preg} 
\end{align}
where $K$ represents scaling factor. The scaling factor is required at the ST to hold $\e{}{\pp}$ at $\ite$. It is defined as
\begin{align}
K = \frac{1}{ \e{\gp, \prcvd}{\frac{\gp \ap}{\prcvd}}},% = \frac{1}{\e{}{\frac{\gs \ap}{\pp}}} 
\label{eq:sca_gen} 
\end{align} 
where, $\e{\gp, \prcvd}{\cdot}$ represents the expectation over $\gp$ and $\prcvd$.
Therefore to implement power control according to (\ref{eq:preg}), it requires the knowledge of probability density function (pdf) $\dK$ and the path loss $\ap$\footnote{For a practical implementation, $\ap$ needs to be evaluated by the ST before power control scheme is applied to the system. 
The path loss $\ap$ in (\ref{eq:sca_gen}) is determined as  
\begin{equation*} 
\ap = \frac{\e{}{\prcvd} - \npp}{\ptran}.
\label{eq:pl}
\end{equation*}
}.
By considering the expression $\gp \ap /\prcvd $ in the denominator of (\ref{eq:sca_gen}) and $\pp = \ite K \ap \gp / \prcvd$ in (\ref{eq:IT}), it is clear that the $\dK$ can be evaluated from $\dpp$. In this way, $K$ can be determined at the ST.
%From (\ref{eq:preg}) $\preg$ depends on $\prcvd$ for a given frame. $\prcvd$ is however estimated from the samples $N(=\tau f\sub{s})$, the estimation of $\prcvd$ involves error. This error is propagated by $P\sub{reg} $ to the performance parameters $P\sub{p}, R$

Given $\tau$ is utilized for $\prcvd$ estimation, the throughput at the SR is given by  
\begin{equation}
\rs = \frac{T - \tau}{T} \log_2 \left(1 + \frac{\gs \as \preg }{\nps} \right). 
\label{eq:Thr_pm}
\end{equation}
It will be clear later in this section that small $\tau$ results in large variations for the $\prcvd$ and vice versa. According to (\ref{eq:preg}), this induces variations in $\preg$. These variations causes $\pp$ to deviate from $\ite$. If not considered in the model, these variations may affect the performance of the system. To deal with this, we define probability of confidence $\pc$ and accuracy $\mu$ that captures the variations of $\pp$ at the PR. As desired by the regulatory bodies, it is important for the system to restrain $\pc$ above a certain desired level $\pcd$ for a given $\mu$. Hence, the probability of confidence constraint is defined as
\begin{equation}
\pc \ge \pcd.
\label{eq:pcc} 
\end{equation}
Clearly, there exits a tradeoff that involves maximizing the expected throughput at the SR subject to a probability of confidence constraint given by 
\begin{align}
\maxi_{\tau}  & \text{      } \e{\gs, \preg}{\rs(\tau)} 
 \label{eq:sys} \\
\text{s.t.} & \text{ } \pc(\tau \fsam, \mu) = \p \left( \frac{\left| \pp - \e{}{\pp} \right|}{\e{}{\pp}} < \mu \right) \ge \pcd, \nonumber  
%\quad 	    & \text{ } \p \left(\frac{\left|\widehat{R} -\e{}{\widehat{R}} \right|}{\e{}{\widehat{R}}} < \mu\sub{R} \right)  \ge \epsilon\sub{R}. \nonumber 
\end{align}
%where $\e{\gs, \preg}{\cdot}$ represents the expectation over $\preg$ at ST and the channel gain $\gs$. 
The probability of confidence is given by 
\begin{align}
\pc = \fpp\left( {(1 + \mu) \e{}{\pp}}\right)  - \fpp\left({(1 - \mu) \e{}{\pp}} \right). \label{eq:pc} 
\end{align}
Now, due to introduction of $K$ in $\preg$, cf. (\ref{eq:preg}) holds the variations in $\pp$ across $\ite$, hence $\e{}{\pp} = 10^{\left( \ite/10 \right)}$. 
In that sense, $\pc$ truly captures the variations of $\pp$ across $\ite$.  
Furthermore, the throughput depicted from (\ref{eq:Thr_cm}) for the conventional model overestimates the throughput of the US. This overestimation in the throughput is evaluated as $\beta$ that depicts the difference between the $\ers$ obtained from the models. 
It is evident that for analyzing the tradeoff depicted in (\ref{eq:sys}), first it is important to characterize the pdfs $\dpp$ and $\drs$. %Now, $\dpp$ and $\drs$ are linked to pdfs $\dpreg$ through (\ref{eq:IT}) and (\ref{eq:Thr_pm}). However, $\preg$ is related to $\prcvd$ via (\ref{eq:preg}).

\subsection{Path loss channel}
To simplify the analysis, we consider a case where the transmitted signals are subject to path loss only, that is, the small scale channel gains correspond to $\gs = \gp = 1$. This way, we first consider the variations in $\pp$ due to the presence of noise in the system. 

\subsubsection*{Characterization of performance parameters}
 
Considering (\ref{eq:prcvd}), $\prcvd$ follows a non-central chi-squared distribution $\mathcal{X'}^2(N \gamma, N)$, where $\gamma = \ap \ptran/\nps$ denotes the received SNR \cite{Urkowitz}. 
%\begin{align}
%\dprcvd(x) =& \frac{N}{2\npp} e^{- \frac{N}{2 \npp}\left( x + \ap \ptran \right)} \left( \frac{x}{\ap \ptran}   \right)^{\frac{N}{4} - \frac{1}{2}} \label{eq:drcvd_pl} \\
%\quad & I_{\frac{N}{2}  - 1}\left( \frac{N}{\npp} \sqrt{x \ap \ptran}  \right), \nonumber
%\end{align}
%where $I_{\frac{N}{2}  - 1}(\cdot)$ represents the Bessel function of order $\frac{N}{2} - 1$ \cite{grad}.

According to (\ref{eq:preg}), $\preg$ follows an inverse non-central chi-squared distribution. The pdf for $\preg$  is given by  
\begin{align}
\dpreg(x) =& \frac{N K \ite}{2\npp x^2} e^{- \frac{N}{2 \npp}\left( \frac{K  \ite}{x} + \ap \ptran \right)} \left( \frac{K \ite}{x \ap \ptran}   \right)^{\frac{N}{4} - \frac{1}{2}} \label{eq:dpreg_pl} \\
\quad & I_{\frac{N}{2}  - 1}\left( \frac{N}{\npp} \sqrt{\frac{ K \ite \ap \ptran}{x}}  \right), \nonumber
%\fpreg(x) = \q{\frac{N}{2} - 1}{\sqrt{\frac{N \ptran \ap}{\npp}},\sqrt{\frac{N \cdot x}{\npp}} }, 
\end{align}
%where $\q{\frac{N}{2} - 1}{\cdot, \cdot}$ is the Marcum-Q function \cite{grad}.
where $I_{\frac{N}{2}  - 1}(\cdot)$ represents the Bessel function of order $\frac{N}{2} - 1$ \cite{grad}.

Following the pdf $\dpreg$ and the relation (\ref{eq:preg}), the pdf of $\pp$ is given by
\begin{align}
\dpp(x) =& \frac{\ap N K \ite}{2\npp x^2} e^{- \frac{N \ap}{2 \npp}\left( \frac{K  \ite}{x} + \ptran \right)} \left( \frac{K \ite}{x \ptran}   \right)^{\frac{N}{4} - \frac{1}{2}} \label{eq:dpp_pl} \\
\quad & I_{\frac{N}{2}  - 1}\left( \frac{N \ap}{\npp} \sqrt{\frac{ K \ite \ptran}{x}}  \right). \nonumber
%\fpp(x) &= \q{\frac{N}{2} - 1}{\sqrt{\frac{N \ptran \ap}{\npp}},\sqrt{\frac{N \cdot \ap \ite K x}{\npp}} },\label{eq:dpp_pl} \\ 
%\text{where, } K &= \ptran + \frac{\npp}{\ap}. \label{eq:sca_pl}
\end{align}
Based on (\ref{eq:dpp_pl}), the distribution function for $\pp$ is determined as 
\begin{align}
\fpp(x) &= 1 - \q{\frac{N}{2} - 1}{\sqrt{\frac{N \ptran \ap}{\npp}},\sqrt{\frac{N \cdot \ap \ite K x}{\npp}} }.  
\label{eq:dfpp_pl}
\end{align}
where $\q{\frac{N}{2} - 1}{\cdot, \cdot}$ is the Marcum-Q function \cite{grad}.
Substituting the distribution (\ref{eq:dfpp_pl}) in (\ref{eq:pc}) to determine $\pc$ for the path loss channel. 

Based on relation (\ref{eq:Thr_cm}) and $\dpreg$, the pdf for $\rs$ is given as  
\begin{align}
\drs(x) =& \frac{N K \ite \as \log2} {2\npp \nps} \left( \frac{p(x) + 1}{[p(x)]^2}\right) e^{- \frac{N}{2 \npp}\left( \frac{K  \ite \as}{p(x) \nps} + \ap \ptran \right)} \label{eq:drs_pl} \\
\quad & \left( \frac{K \ite \as}{p(x) \ap \ptran \nps}   \right)^{\frac{N}{4} - \frac{1}{2}}  I_{\frac{N}{2}  - 1}\left( \frac{N}{\npp} \sqrt{\frac{ K \ite \ap \ptran \as}{p(x) \nps}}  \right), \nonumber 
%\frs(x) &= \q{\frac{N}{2} - 1}{\sqrt{\frac{N \ptran \ap}{\npp}},\sqrt{\frac{N \cdot \ap \ite K \as (2^x - 1)}{\npp \nps}} },\label{eq:drs_pl}  
\end{align}
where, $p(x) = (2^x - 1)$.   
Consequently, from (\ref{eq:drs_pl}), the expected throughput at the SR for the path loss channel is evaluated as 
\begin{align}
\e{}{\rs(\tau)} = \int\limits_{0}^{\infty} x \drs(x) dx. 
%\e{}{\rs(\tau)} = \frac{T -\tau}{T} \log_2 \left(1 + \frac{\ite K \as}{(\ptran \ap +  \npp) \nps}   \right). 
\label{eq:ers_pl}
\end{align}

\subsubsection*{Noise uncertainty}
The characterization of $\pc$ and $\e{}{\rs}$ based on $\prcvd$ requires the perfect knowledge of $\npp$ at the ST. In this regard, we determine the influence of noise uncertainty on the performance of the system. Following \cite{Tan08}, the noise power $\npp$ can be expressed as a bounded interval $\left[1/\rho \cdot \npp, \rho \cdot \npp \right]$, where $\rho$ amounts the level of noise uncertainty in the system. To simplify the analysis, we represent noise uncertainty as $\npu = 10^{\rho/10}$. Therefore, to characterize the effect of noise uncertainty, we substitute $\npp$ as $\npp \pm \npu$ in (\ref{eq:pc}). This way, we develop the best and worst performance bounds for the US. 
\subsection{Fading channel}
Now, we extend the analysis of the estimation-throughput tradeoff to a fading scenario, which is found more often in practice. Fading causes random variations in the channel. According to the system model, these variations remain constant for a complete frame duration. To cancel the effect of these variations, an ST implements power control for each frame in order to sustain (\ref{eq:IT}). In this section, we investigate the effect of channel estimation error on the performance of the US with fading. 
\subsubsection*{Characterization of performance parameters}
%In order to consider the estimation-throughput tradeoff described in (\ref{eq:sys}), it is required 
For determining the expressions of $\pc$ and $\e{}{\rs}$ for the fading channel, it is required to determine the pdf for $\preg$. Based on this, we obtain the pdfs for $\pp$ and $\rs$. The exact analytical expression of the pdfs for $\preg$ and $\pp$ are given by (\ref{eq:dpreg_f}) and (\ref{eq:dpp_f}), see the top of the next page, where $\Gamma(\cdot)$ denotes the Gamma function, $\Gamma(\cdot, \cdot)$ represents the incomplete Gamma function and ${}_2 F_1(\cdot, \cdot, \cdot, \cdot)$ in (\ref{eq:dpp_f}) is the Hypergeometric function \cite{grad}.   
\begin{figure*}[!t]
\normalsize
\begin{align}
\label{eq:dpreg_f}
\dpreg(x) &= \exp\left( - \frac{\ite N K}{2 \npp x + N \ap \ptran x} \right) (N - 2) N \ap \npp \ap \ptran \left( 2 + \frac{N \ap \ptran}{\npp}  \right)^{\frac{N}{2}} \left( \frac{\ite K}{\ap \ptran x} \right)^{\frac{2 + N}{4} }  \left( \frac{\ite N^2 K \ap \ptran}{\sigma^4\sub{p} x }  \right)^{\frac{1}{2} - \frac{N}{4}} \\ 
\quad & \left( \Gamma\left[\frac{N}{2} - 1 \right] - \Gamma\left[\frac{N}{2} - 1, \frac{\ite N^2 K \ap \ptran}{(4 \sigma^4\sub{p} x + 2 N \npp \ap \ptran x)} \right]  \right) / \left( 2 (2 \npp + N \ap \ptran)^2 x \Gamma\left[ \frac{N}{2} \right]   \right). \nonumber
\end{align}

\begin{align}
\label{eq:dpp_f}
\dpp(x) &= \frac{1}{\ite K \ap} \npp \left( \frac{\ite K}{\ptran x} \right)^{\frac{2 + N}{4}} [h(x)]^{\frac{2 + N}{4}} {}_2 F_1 \left[ \frac{2 + N}{4}, \frac{4 + N}{4}, \frac{N}{2}, 4 h(x) \right],
\text{where } h(x)  = \frac{\ite N^2 K \ap^2 \ptran x}{(\ite N K \ap + 2 \npp x + N \ap \ptran x )^2 } 
\end{align}
% IEEE uses as a separator
\hrulefill
% The spacer can be tweaked to stop underfull vboxes.
\vspace*{4pt}
\end{figure*}

%According to (\ref{eq:sca_gen}), the scaling constant $K$ for the fading case is given by 
%\begin{align}
%K = \int\limits_{0}^{\infty} \frac{\npp + x \ap \ptran}{x \ap} e^{-x} dx
%\label{eq:sca_f}
%\end{align}
With the characterization of $\dpp$, $\pc$ for the fading channel can be evaluated according to (\ref{eq:pc}). Moreover, the expression for $\dpreg$ in (\ref{eq:dpreg_f}) is used to determine the pdf for $\rs$ as
\begin{align}
\drs(x) = \int\limits_{0}^{\infty} \frac{1}{\as \gs} \dpreg\left( \frac{2^x - 1}{\as \gs} \right) e^{-\gs} d\gs
\label{eq:drs_f}
\end{align} 
According to (\ref{eq:ers_pl}), the expression for $\drs$ in (\ref{eq:drs_f}) is used to evaluate the expected throughput $\ers$ for the fading channel. 
%\begin{align}
%\e{}{\rs(\tau)} = \int\limits_{0}^{\infty} x \drs(x) dx 
%\label{eq:ers_f}
%\end{align}
 
%%%%%%%%%%%%%%%%%%%%%%%%%%%%%%%%%%%%%%%%%%%%%%%%%%%%%%%%%%%%%%%%%%%%%%%%%%%%%%%%%%%%%%%%%
\section{Numerical Analysis} \label{sec:num_ana}
%%%%%%%%%%%%%%%%%%%%%%%%%%%%%%%%%%%%%%%%%%%%%%%%%%%%%%%%%%%%%%%%%%%%%%%%%%%%%%%%%%%%%%%%%
In this section, we evaluate the performance of the US for the proposed model. In this regard, we perform simulations to validate the expressions obtained in the previous section. Moreover, the system parameters are selected such that they conform to the scenario described in \figurename~\ref{fig:scenario}. To investigate the performance of the system based on the estimation model, in the analysis, the conventional model is used as a benchmark. 
%\subsection{System parameters}
Unless explicitly mentioned, the following choice of the system parameters is considered for the analysis, $\ptran = \SI{0}{dBm}$, $\ite = -\SI{110}{dBm}$, $\fsam = \SI{1}{MHz}$ $\gamma = \SI{0}{dB}$, $\ap = \SI{100}{dB}, \as = \SI{80}{dB}$, $T = \SI{100}{ms}$, $\mu = 0.025$, $\pcd = 0.95$, $\npp = \nps = \SI{-100}{dBm}$ and $\npu = \pm \SI{3}{dB}$. 

\subsection{Path loss channel}
%\subsubsection{Estimation throughput tradeoff}
Firstly, the analysis of the estimation throughput tradeoff based on (\ref{eq:sys}) is performed. \figurename~\ref{fig:ET_pl} demonstrates the variation of the performance parameters $\ers$ and $\pc$ with $\tau$. It is evident that the expected throughput decreases linearly with $\tau$, while the throughput according to the conventional model described in (\ref{eq:Thr_cm}) remains constant. The slope of $\ers$ depends on the choice of system parameters. On the other side, $\pc$ increases with $\tau$ according to (\ref{eq:pc}). The horizontal dashed line represents the $\pc$ constraint, cf. (\ref{eq:pcc}). Solving (\ref{eq:sys}), to obtain the maximum $\ers$. Hence, according to (\ref{eq:sys}), it is required to determine $\tau$ such that $\pc = \pcd$. This corresponds to the projection of the point $\pc = \pcd$ on the curve $\ers$. The projection is illustrated through a vertical dashed line, cf. \figurename~\ref{fig:ET_pl}. Hence, the circle on the curve $\ers$ denotes the maximum throughput with $\npu = \SI{0}{dB}$ under the considered system parameters. 

Finally, we include noise uncertainty $\npu$ in the estimation throughput analysis and determine the performance bounds for the proposed model. Clearly, $\npu = -\SI{3}{dB}$ signifies a better SNR that improves $\pc$ of the system, cf. \figurename~{\ref{fig:ET_pl}. With $\npu = -\SI{3}{dB}$, the probability of confidence constraint ($\pc = \pcd$) is fulfilled at a lower $\tau$, leading to a larger $\ers$. Consequently, $\npu = -\SI{3}{dB}$ attains the best performance bound for the US. On similar basis, $\npu = \SI{3}{dB}$ determines the worst performance bound.  

\begin{figure}[!t]
% This file is generated by the MATLAB m-file laprint.m. It can be included
% into LaTeX documents using the packages graphicx, color and psfrag.
% It is accompanied by a postscript file. A sample LaTeX file is:
%    \documentclass{article}\usepackage{graphicx,color,psfrag}
%    \begin{document}\input{fig_thr_est_time_tradeoff_AWGN}\end{document}
% See http://www.mathworks.de/matlabcentral/fileexchange/loadFile.do?objectId=4638
% for recent versions of laprint.m.
%
% created by:           LaPrint version 3.16 (13.9.2004)
% created on:           05-Feb-2015 04:59:16
% eps bounding box:     14 cm x 10.5 cm
% comment:              
%
%\begin{psfrags}%
%\psfragscanon%
%
% text strings:
\psfrag{s08}[b][b]{\fontsize{9}{13.5}\fontseries{m}\mathversion{normal}\fontshape{n}\selectfont \color[rgb]{0,0,1}\setlength{\tabcolsep}{0pt}\begin{tabular}{c}$\ers$ = [bits/sec/Hz]\end{tabular}}%
\psfrag{s09}[t][t]{\fontsize{9}{13.5}\fontseries{m}\mathversion{normal}\fontshape{n}\selectfont \color[rgb]{0,0,0}\setlength{\tabcolsep}{0pt}\begin{tabular}{c}$\tau$ = [ms]\end{tabular}}%
\psfrag{s13}[][]{\fontsize{10}{15}\fontseries{m}\mathversion{normal}\fontshape{n}\selectfont \color[rgb]{0,0,0}\setlength{\tabcolsep}{0pt}\begin{tabular}{c} \end{tabular}}%
\psfrag{s14}[][]{\fontsize{10}{15}\fontseries{m}\mathversion{normal}\fontshape{n}\selectfont \color[rgb]{0,0,0}\setlength{\tabcolsep}{0pt}\begin{tabular}{c} \end{tabular}}%
\psfrag{s16}[t][t]{\fontsize{9}{13.5}\fontseries{m}\mathversion{normal}\fontshape{n}\selectfont \color[rgb]{0,0.5,0}\setlength{\tabcolsep}{0pt}\begin{tabular}{c}$\pc$\end{tabular}}%
\psfrag{s17}[t][t]{\fontsize{9}{13.5}\fontseries{m}\mathversion{normal}\fontshape{n}\selectfont \color[rgb]{0,0,0}\setlength{\tabcolsep}{0pt}\begin{tabular}{c}$\tau$ = [ms]\end{tabular}}%
\psfrag{s18}[l][l]{\fontsize{9}{13.5}\fontseries{m}\mathversion{normal}\fontshape{n}\selectfont \color[rgb]{0,0,0}sim}%
\psfrag{s19}[l][l]{\fontsize{9}{13.5}\fontseries{m}\mathversion{normal}\fontshape{n}\selectfont \color[rgb]{0,0,0}(6)}%
\psfrag{s20}[l][l]{\fontsize{9}{13.5}\fontseries{m}\mathversion{normal}\fontshape{n}\selectfont \color[rgb]{0,0,0}$\pc$, $\npu$ = 0 dB}%
\psfrag{s21}[l][l]{\fontsize{9}{13.5}\fontseries{m}\mathversion{normal}\fontshape{n}\selectfont \color[rgb]{0,0,0}$\pc, \npu = \pm 3$ dB}%
\psfrag{s22}[l][l]{\fontsize{9}{13.5}\fontseries{m}\mathversion{normal}\fontshape{n}\selectfont \color[rgb]{0,0,0}max $\tau, \npu = 0$ dB}%
\psfrag{s23}[l][l]{\fontsize{9}{13.5}\fontseries{m}\mathversion{normal}\fontshape{n}\selectfont \color[rgb]{0,0,0}max $\tau, \npu = \pm 3$ dB}%
\psfrag{s24}[l][l]{\fontsize{9}{13.5}\fontseries{m}\mathversion{normal}\fontshape{n}\selectfont \color[rgb]{0,0,0}sim}%
%
% axes font properties:
\fontsize{9}{13.5}\fontseries{m}\mathversion{normal}%
\fontshape{n}\selectfont%
%
% xticklabels:
\psfrag{x01}[t][t]{2}%
\psfrag{x02}[t][t]{4}%
\psfrag{x03}[t][t]{6}%
\psfrag{x04}[t][t]{8}%
\psfrag{x05}[t][t]{10}%
\psfrag{x06}[t][t]{12}%
\psfrag{x07}[t][t]{14}%
\psfrag{x08}[t][t]{16}%
\psfrag{x09}[t][t]{18}%
\psfrag{x10}[t][t]{20}%
\psfrag{x11}[t][t]{2}%
\psfrag{x12}[t][t]{4}%
\psfrag{x13}[t][t]{6}%
\psfrag{x14}[t][t]{8}%
\psfrag{x15}[t][t]{10}%
\psfrag{x16}[t][t]{12}%
\psfrag{x17}[t][t]{14}%
\psfrag{x18}[t][t]{16}%
\psfrag{x19}[t][t]{18}%
\psfrag{x20}[t][t]{20}%
%
% yticklabels:
\psfrag{v02}[l][l]{0.5}%
\psfrag{v03}[l][l]{0.6}%
\psfrag{v04}[l][l]{0.7}%
\psfrag{v05}[l][l]{0.8}%
\psfrag{v06}[l][l]{0.9}%
\psfrag{v01}[l][l]{1}%
\psfrag{v07}[r][r]{2.77}%
\psfrag{v08}[r][r]{2.89}%
\psfrag{v09}[r][r]{3}%
\psfrag{v10}[r][r]{3.12}%
\psfrag{v11}[r][r]{3.24}%
\psfrag{v12}[r][r]{3.36}%
%
% Figure:
%\resizebox{7cm}{!}{\includegraphics{fig_thr_est_time_tradeoff_AWGN.eps}}%
%\end{psfrags}%
%
% End fig_thr_est_time_tradeoff_AWGN.tex

\centering
\begin{tikzpicture}[scale=1]
\node[anchor=south west,inner sep=0] (image) at (0,0)
{
        \includegraphics[width=\columnwidth]{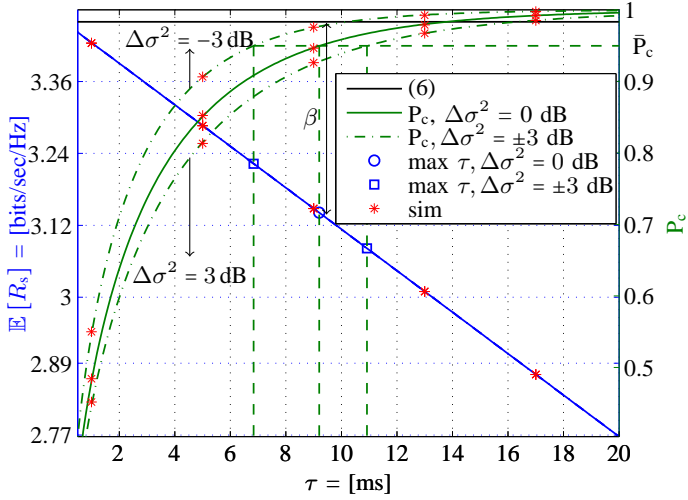}
};
\begin{scope}[x={(image.south east)},y={(image.north west)}]

% Select curves depending upon theta   
%\draw (0.34,0.88) arc(-130:130:0.70mm and 1.20mm)  node[left, font=\small] {$\npu$ = $\SI{-3}{dB}$};
%\draw (0.25,0.62) arc(-130:130:0.70mm and 1.20mm)  node[below, font=\small] {$\npu$ = $\SI{3}{dB}$};
\draw[black,->] (0.26,0.66) --  node[below=6.2mm, font=\small] {$\npu = \SI{3}{dB}$} (0.26,0.46);
\draw[black,->] (0.26,0.80) --  node[above=1.7mm, font=\small] {$\npu = \SI{-3}{dB}$} (0.26,0.88);

\draw(0.94,0.885) node[font=\small]{$\pcd$};
\draw[black,<->] (0.465,0.545) --  node[left=0.0mm, font=\small] {$\beta$} (0.465,0.932);

%\draw[help lines,xstep=.1,ystep=.1] (0,0) grid (1,1);
%\foreach \x in {0,1,...,9} { \node [anchor=north] at (\x/10,0) {0.\x}; }
%\foreach \y in {0,1,...,9} { \node [anchor=east] at (0,\y/10) {0.\y}; }
\end{scope}
\end{tikzpicture}
\caption{An illustration of estimation-throughput tradeoff for the path loss channel with $\gamma = \SI{0}{dB}$. $\beta$ represents the least performance loss for $\gamma = \SI{0}{dB}$ where the constraint for $\pc = \pcd$ is sustained.}
\label{fig:ET_pl}
\vspace{-5mm}
\end{figure}

%\begin{figure}[!t]
%        \includegraphics[trim=0.6cm 0.4cm 0.2cm 1.2cm,clip=true,width=\columnwidth]{figures/fig_thr_est_time_tradeoff_AWGN}
%\caption{An illustration of estimation-throughput tradeoff for the path loss channel with $\gamma = \SI{0}{dB}$.}
%\label{fig:ET_pl}
%\end{figure}

In the previous analysis, we obtained the maximum throughput for the US while operating the system at $\pc = \pcd$. Hence for further analysis, the system is analyzed considering the maximum throughput. Additionally, we use the analytical expressions for plotting the curves.   
%\subsubsection{Optimum throughput vs SNR at ST} 

Next, it is interesting to consider the influence of received SNR on the performance of the system with different $\pcd$. From \figurename~\ref{fig:TvSNR_pl}, it is observed that $\ers$ is more sensitive to $\gamma$ for $\gamma \le \SI{5}{dB}$. This becomes clear by understanding the influence of $\gamma$ on $\pc$. That is, with increasing $\gamma$ the curvature for the curve $\pc$ increases. However, its projection representing $\ers$ lies on an elevated line with a certain slope. This explains the shape of curve obtained for $\ers$. Moreover, as $\pcd$ increases, $\ers$ becomes sensitive towards the variation in $\gamma$, cf. \figurename~\ref{fig:TvSNR_pl}. This is due to the fact that by reducing $\pcd$, the system tends towards the larger curvature. %Finally, for $\pcd = 0.95$, the effect of $\npu$ on the performance of the system is demonstrated. The noise uncertainty has a negligible influence on $\ers$ at $\gamma < -\SI{10}{dB}$, cf. \figurename~\ref{fig:TvSNR_pl}, however the effect becomes significant for $\gamma \in [-10, 15]\SI{}{dB}$.   

\begin{figure}[!t]
% This file is generated by the MATLAB m-file laprint.m. It can be included
% into LaTeX documents using the packages graphicx, color and psfrag.
% It is accompanied by a postscript file. A sample LaTeX file is:
%    \documentclass{article}\usepackage{graphicx,color,psfrag}
%    \begin{document}\input{fig_opt_thr_vs_snr_AWGN}\end{document}
% See http://www.mathworks.de/matlabcentral/fileexchange/loadFile.do?objectId=4638
% for recent versions of laprint.m.
%
% created by:           LaPrint version 3.16 (13.9.2004)
% created on:           05-Feb-2015 04:53:36
% eps bounding box:     14 cm x 10.5 cm
% comment:              
%
%\begin{psfrags}%
%\psfragscanon%
%
% text strings:
\psfrag{s01}[b][b]{\fontsize{9}{13.5}\fontseries{m}\mathversion{normal}\fontshape{n}\selectfont \color[rgb]{0,0,0}\setlength{\tabcolsep}{0pt}\begin{tabular}{c}$\ers$ = [bits/sec/Hz]\end{tabular}}%
\psfrag{s02}[t][t]{\fontsize{9}{13.5}\fontseries{m}\mathversion{normal}\fontshape{n}\selectfont \color[rgb]{0,0,0}\setlength{\tabcolsep}{0pt}\begin{tabular}{c}$\gamma$ = [dB]\end{tabular}}%
\psfrag{s06}[][]{\fontsize{10}{15}\fontseries{m}\mathversion{normal}\fontshape{n}\selectfont \color[rgb]{0,0,0}\setlength{\tabcolsep}{0pt}\begin{tabular}{c} \end{tabular}}%
\psfrag{s07}[][]{\fontsize{10}{15}\fontseries{m}\mathversion{normal}\fontshape{n}\selectfont \color[rgb]{0,0,0}\setlength{\tabcolsep}{0pt}\begin{tabular}{c} \end{tabular}}%
\psfrag{s08}[l][l]{\fontsize{9}{13.5}\fontseries{m}\mathversion{normal}\fontshape{n}\selectfont \color[rgb]{0,0,0}$\pcd = 0.97$}%
\psfrag{s09}[l][l]{\fontsize{9}{13.5}\fontseries{m}\mathversion{normal}\fontshape{n}\selectfont \color[rgb]{0,0,0}(6)}%
\psfrag{s10}[l][l]{\fontsize{9}{13.5}\fontseries{m}\mathversion{normal}\fontshape{n}\selectfont \color[rgb]{0,0,0}$\pcd = 0.92$}%
\psfrag{s11}[l][l]{\fontsize{9}{13.5}\fontseries{m}\mathversion{normal}\fontshape{n}\selectfont \color[rgb]{0,0,0}$\pcd = 0.95$}%
\psfrag{s12}[l][l]{\fontsize{9}{13.5}\fontseries{m}\mathversion{normal}\fontshape{n}\selectfont \color[rgb]{0,0,0}$\pcd = 0.97$}%
%
% axes font properties:
\fontsize{9}{13.5}\fontseries{m}\mathversion{normal}%
\fontshape{n}\selectfont%
%
% xticklabels:
\psfrag{x01}[t][t]{-15}%
\psfrag{x02}[t][t]{-10}%
\psfrag{x03}[t][t]{-5}%
\psfrag{x04}[t][t]{0}%
\psfrag{x05}[t][t]{5}%
\psfrag{x06}[t][t]{10}%
%
% yticklabels:
\psfrag{v01}[r][r]{1}%
\psfrag{v02}[r][r]{2}%
\psfrag{v03}[r][r]{3}%
\psfrag{v04}[r][r]{4}%
\psfrag{v05}[r][r]{5}%
\psfrag{v06}[r][r]{6}%
\psfrag{v07}[r][r]{7}%
\psfrag{v08}[r][r]{8}%
%
% Figure:
%\resizebox{7cm}{!}{\includegraphics{fig_opt_thr_vs_snr_AWGN.eps}}%
%\end{psfrags}%
%
% End fig_opt_thr_vs_snr_AWGN.tex

\centering
\begin{tikzpicture}[scale=1]
\node[anchor=south west,inner sep=0] (image) at (0,0)
{
	\includegraphics[width= 0.95 \columnwidth]{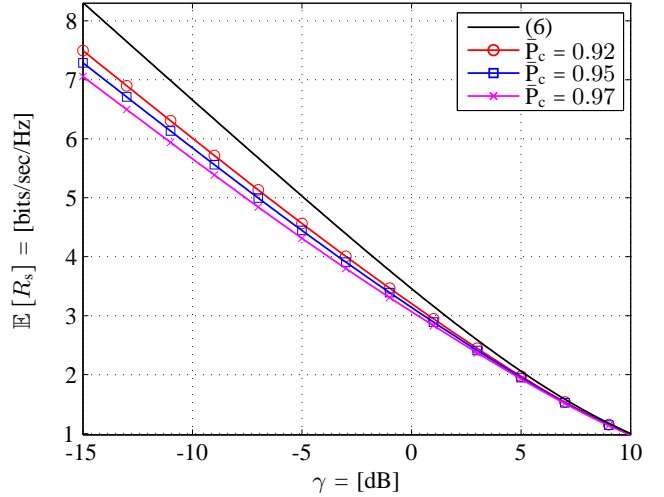}
};
\begin{scope}[x={(image.south east)},y={(image.north west)}]

% Select curves depending upon theta   
%\draw (0.2,0.405) arc(-130:130:0.70mm and 1.20mm)  node[above, font=\small] {$\pcd = 0.92$};
%\draw (0.2,0.26) arc(-130:130:0.70mm and 1.20mm)  node[above, font=\small] {$\pcd = 0.95$};
%\draw (0.35,0.125) arc(-130:130:0.70mm and 1.20mm)  node[above, font=\small] {$\pcd = 0.97$};
%\draw (0.7,0.875) arc(-130:130:0.70mm and 1.20mm)  node[left, font=\small] {$\npu$ = $\SI{-3}{dB}$};
%\draw (0.7,0.645) arc(-130:130:0.70mm and 1.20mm)  node[right=1.00mm, font=\small] {$\npu$ = $\SI{+3}{dB}$};
%\draw[black,->] (0.2,0.52) --  node[above=4.2mm,right=2.1mm, font=\tiny] {$\alpha$} (0.27,0.62);
%\draw[black,->] (0.815,0.22) --  node[above=4.2mm,right=2.1mm, font=\tiny] {$\alpha$} (0.745,0.12);

% Select curves depending upon theta   
%\draw (0.7,0.875) arc(-130:130:0.70mm and 1.20mm)  node[left, font=\small] {$\npu$ = $\SI{-3}{dB}$};
%\draw[help lines,xstep=.1,ystep=.1] (0,0) grid (1,1);
%\foreach \x in {0,1,...,9} { \node [anchor=north] at (\x/10,0) {0.\x}; }
%\foreach \y in {0,1,...,9} { \node [anchor=east] at (0,\y/10) {0.\y}; }
\end{scope}
\end{tikzpicture}
\caption{An illustration of the dependency of the received SNR at ST on the maximum throughput at the SR for different values of probability of confidence constraint $\pcd \in \{0.92, 0.95, 0.97\}$. The variation in $\ap \in [85,110]$ \SI{}{dB} translates to the variation in $\gamma$.}
\label{fig:TvSNR_pl}
%\vspace{-5mm}
\end{figure}
%\begin{figure}[!t]
%\makeatletter
%        \includegraphics[trim=0.6cm 0.4cm 0.2cm 1.2cm,clip=true,width=\columnwidth]{figures/fig_opt_thr_vs_snr_AWGN}
%\caption{An illustration of the dependency of the received SNR at ST on the maximum throughput at SR for different values of confidence probability $\pcd = [0.92, 0.95, 0.97]$.}
%\label{fig:TvSNR_pl}
%\end{figure}

%\subsubsection{Optimum throughput vs accuracy}
\begin{figure}[!t]
% This file is generated by the MATLAB m-file laprint.m. It can be included
% into LaTeX documents using the packages graphicx, color and psfrag.
% It is accompanied by a postscript file. A sample LaTeX file is:
%    \documentclass{article}\usepackage{graphicx,color,psfrag}
%    \begin{document}\input{fig_opt_thr_vs_acc_AWGN}\end{document}
% See http://www.mathworks.de/matlabcentral/fileexchange/loadFile.do?objectId=4638
% for recent versions of laprint.m.
%
% created by:           LaPrint version 3.16 (13.9.2004)
% created on:           05-Feb-2015 04:54:50
% eps bounding box:     14 cm x 10.5 cm
% comment:              
%
%\begin{psfrags}%
%\psfragscanon%
%
% text strings:
\psfrag{s01}[b][b]{\fontsize{9}{13.5}\fontseries{m}\mathversion{normal}\fontshape{n}\selectfont \color[rgb]{0,0,0}\setlength{\tabcolsep}{0pt}\begin{tabular}{c}$\ers$ = [bits/sec/Hz]\end{tabular}}%
\psfrag{s02}[t][t]{\fontsize{9}{13.5}\fontseries{m}\mathversion{normal}\fontshape{n}\selectfont \color[rgb]{0,0,0}\setlength{\tabcolsep}{0pt}\begin{tabular}{c}Accuracy ($\mu$)\end{tabular}}%
\psfrag{s06}[][]{\fontsize{10}{15}\fontseries{m}\mathversion{normal}\fontshape{n}\selectfont \color[rgb]{0,0,0}\setlength{\tabcolsep}{0pt}\begin{tabular}{c} \end{tabular}}%
\psfrag{s07}[][]{\fontsize{10}{15}\fontseries{m}\mathversion{normal}\fontshape{n}\selectfont \color[rgb]{0,0,0}\setlength{\tabcolsep}{0pt}\begin{tabular}{c} \end{tabular}}%
\psfrag{s08}[l][l]{\fontsize{9}{13.5}\fontseries{m}\mathversion{normal}\fontshape{n}\selectfont \color[rgb]{0,0,0}$\pcd = 0.97$}%
\psfrag{s09}[l][l]{\fontsize{9}{13.5}\fontseries{m}\mathversion{normal}\fontshape{n}\selectfont \color[rgb]{0,0,0}(6)}%
\psfrag{s10}[l][l]{\fontsize{9}{13.5}\fontseries{m}\mathversion{normal}\fontshape{n}\selectfont \color[rgb]{0,0,0}$\pcd = 0.92$}%
\psfrag{s11}[l][l]{\fontsize{9}{13.5}\fontseries{m}\mathversion{normal}\fontshape{n}\selectfont \color[rgb]{0,0,0}$\pcd = 0.95$}%
\psfrag{s12}[l][l]{\fontsize{9}{13.5}\fontseries{m}\mathversion{normal}\fontshape{n}\selectfont \color[rgb]{0,0,0}$\pcd = 0.97$}%
%
% axes font properties:
\fontsize{9}{13.5}\fontseries{m}\mathversion{normal}%
\fontshape{n}\selectfont%
%
% xticklabels:
\psfrag{x01}[t][t]{0.015}%
\psfrag{x02}[t][t]{0.02}%
\psfrag{x03}[t][t]{0.025}%
\psfrag{x04}[t][t]{0.03}%
\psfrag{x05}[t][t]{0.035}%
\psfrag{x06}[t][t]{0.04}%
\psfrag{x07}[t][t]{0.045}%
\psfrag{x08}[t][t]{0.05}%
\psfrag{x09}[t][t]{0.055}%
%
% yticklabels:
\psfrag{v01}[r][r]{1.5}%
\psfrag{v02}[r][r]{2}%
\psfrag{v03}[r][r]{2.5}%
\psfrag{v04}[r][r]{3}%
\psfrag{v05}[r][r]{3.5}%
%
% Figure:
%\resizebox{7cm}{!}{\includegraphics{fig_opt_thr_vs_acc_AWGN.eps}}%
%\end{psfrags}%
%
% End fig_opt_thr_vs_acc_AWGN.tex

\centering
\begin{tikzpicture}[scale=1]
\node[anchor=south west,inner sep=0] (image) at (0,0)
{
        \includegraphics[width=0.95 \columnwidth]{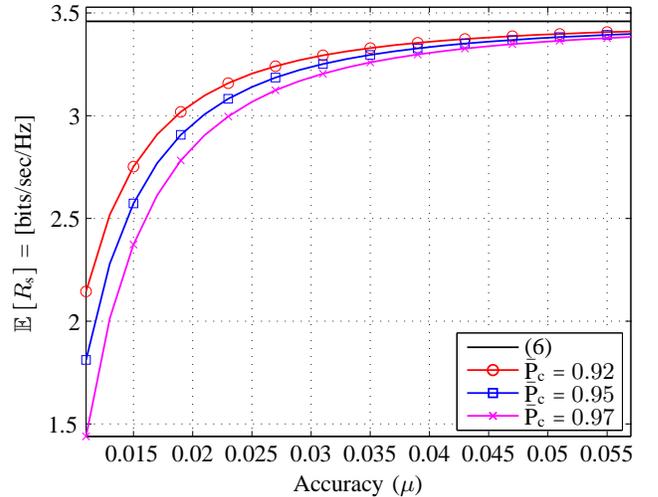}
};
\begin{scope}[x={(image.south east)},y={(image.north west)}]

% Select curves depending upon theta   
%\draw (0.7,0.875) arc(-130:130:0.70mm and 1.20mm)  node[left, font=\small] {$\npu$ = $\SI{-3}{dB}$};
%\draw[black,->] (0.3,0.775) --  node[above=-6.0mm,right=0mm, font=\small] {$\pcd = 0.97$} (0.4,0.675);
%\draw[black,->] (0.2,0.66) --  node[above=-6.0mm,right=0mm, font=\small] {$\pcd = 0.95$} (0.3,0.56);
%\draw[black,->] (0.1255,0.495) --  node[above=-6.0mm,right=0mm, font=\small] {$\pcd = 0.92$} (0.2255,0.395);

%\draw[black,->] (0.18,0.55) --  node[above=-15.0mm,right=0mm, font=\small] {$\pcd = 0.92, \npu = \SI{+3}{dB}$} (0.43,0.3);
%\draw[black,->] (0.23,0.77) --  node[above=-15.0mm,right=0mm, font=\small] {$\pcd = 0.92, \npu = \SI{-3}{dB}$} (0.48,0.52);
%\draw[help lines,xstep=.1,ystep=.1] (0,0) grid (1,1);
%\foreach \x in {0,1,...,9} { \node [anchor=north] at (\x/10,0) {0.\x}; }
%\foreach \y in {0,1,...,9} { \node [anchor=east] at (0,\y/10) {0.\y}; }
\end{scope}
\end{tikzpicture}
\caption{A variation of the maximum throughput at the SR with estimation accuracy $\mu$ for different probability of confidence constraint $\pcd \in \{0.92, 0.95, 0.97\}$ with fixed $\gamma = \SI{0}{dB}$.}
\label{fig:TvA_pl}
\vspace{-5mm}
\end{figure}
%\begin{figure}[!t]
%\makeatletter
%        \includegraphics[trim=0.6cm 0.4cm 0.2cm 1.2cm,clip=true,width=\columnwidth]{figures/fig_opt_thr_vs_acc_AWGN}
%\caption{A variation of the maximum throughput at SR with estimation accuracy $\mu$ for different confidence probability $\pcd = [0.92, 0.95, 0.97]$ with fixed $\gamma = \SI{0}{dB}$.}
%\label{fig:TvA_pl}
%\end{figure}
For the final analysis concerning the path loss channel, we consider the effect of the estimation accuracy on $\ers$ for fixed $\gamma = \SI{0}{dB}$ and different $\pcd \in \{0.92,0.95,0.97\}$. It is visible from \figurename~\ref{fig:TvA_pl} that $\ers$ decays exponentially with $\mu$. In order to capture this effect, we consider the characterization of $\pc$ in (\ref{eq:pc}). It is evident that $\pc$ is related to $\mu$ via Marcum-Q function considered in (\ref{eq:dpp_pl}). Consequently, for a fixed $\pc$, the variation in $\mu$ is compensated by $\tau$. Following (\ref{eq:drs_pl}), this variation is mapped into $\ers$. %Moreover, a large variation in $\ers$ is witnessed against different $\pcd$ at smaller $\mu$, refer \figurename~\ref{fig:TvA_pl}. The variation is negligible for bigger $\mu$. %In addition to that, for $\pcd = 0.95$ is plotted. Likewise $\pcd$, $\npu$ has greater influence on the performance at lower $\mu$.   

\subsection{Fading channel}
In this section, we extend our analysis to a channel that involves fading. The curves for $\pc$ and $\ers$ were generated through simulation. However for $\pc$, we validate the simulation results through the analytical expression, cf. (\ref{eq:pc}).  
%\subsubsection{Estimation throughput tradeoff}

\begin{figure}[!t]
% This file is generated by the MATLAB m-file laprint.m. It can be included
% into LaTeX documents using the packages graphicx, color and psfrag.
% It is accompanied by a postscript file. A sample LaTeX file is:
%    \documentclass{article}\usepackage{graphicx,color,psfrag}
%    \begin{document}\input{fig_thr_est_time_tradeoff_fading}\end{document}
% See http://www.mathworks.de/matlabcentral/fileexchange/loadFile.do?objectId=4638
% for recent versions of laprint.m.
%
% created by:           LaPrint version 3.16 (13.9.2004)
% created on:           05-Feb-2015 08:24:42
% eps bounding box:     14 cm x 10.5 cm
% comment:              
%
%\begin{psfrags}%
%\psfragscanon%
%
% text strings:
\psfrag{s08}[b][b]{\fontsize{9}{13.5}\fontseries{m}\mathversion{normal}\fontshape{n}\selectfont \color[rgb]{0,0,1}\setlength{\tabcolsep}{0pt}\begin{tabular}{c}$\ers$ = [bits/sec/Hz]\end{tabular}}%
\psfrag{s09}[t][t]{\fontsize{9}{13.5}\fontseries{m}\mathversion{normal}\fontshape{n}\selectfont \color[rgb]{0,0,0}\setlength{\tabcolsep}{0pt}\begin{tabular}{c}$\tau$ = [ms]\end{tabular}}%
\psfrag{s13}[][]{\fontsize{10}{15}\fontseries{m}\mathversion{normal}\fontshape{n}\selectfont \color[rgb]{0,0,0}\setlength{\tabcolsep}{0pt}\begin{tabular}{c} \end{tabular}}%
\psfrag{s14}[][]{\fontsize{10}{15}\fontseries{m}\mathversion{normal}\fontshape{n}\selectfont \color[rgb]{0,0,0}\setlength{\tabcolsep}{0pt}\begin{tabular}{c} \end{tabular}}%
\psfrag{s16}[t][t]{\fontsize{9}{13.5}\fontseries{m}\mathversion{normal}\fontshape{n}\selectfont \color[rgb]{0,0.5,0}\setlength{\tabcolsep}{0pt}\begin{tabular}{c}$\pc$\end{tabular}}%
\psfrag{s17}[t][t]{\fontsize{9}{13.5}\fontseries{m}\mathversion{normal}\fontshape{n}\selectfont \color[rgb]{0,0,0}\setlength{\tabcolsep}{0pt}\begin{tabular}{c}$\tau$ = [ms]\end{tabular}}%
\psfrag{s18}[l][l]{\fontsize{9}{13.5}\fontseries{m}\mathversion{normal}\fontshape{n}\selectfont \color[rgb]{0,0,0}theory}%
\psfrag{s19}[l][l]{\fontsize{9}{13.5}\fontseries{m}\mathversion{normal}\fontshape{n}\selectfont \color[rgb]{0,0,0}$\ers$}%
\psfrag{s20}[l][l]{\fontsize{9}{13.5}\fontseries{m}\mathversion{normal}\fontshape{n}\selectfont \color[rgb]{0,0,0}$\pc$}%
\psfrag{s21}[l][l]{\fontsize{9}{13.5}\fontseries{m}\mathversion{normal}\fontshape{n}\selectfont \color[rgb]{0,0,0}(6)}%
\psfrag{s22}[l][l]{\fontsize{9}{13.5}\fontseries{m}\mathversion{normal}\fontshape{n}\selectfont \color[rgb]{0,0,0}theory}%
%
% axes font properties:
\fontsize{9}{13.5}\fontseries{m}\mathversion{normal}%
\fontshape{n}\selectfont%
%
% xticklabels:
\psfrag{x01}[t][t]{$10^{-2}$}%
\psfrag{x02}[t][t]{$10^{-1}$}%
\psfrag{x03}[t][t]{$10^{0}$}%
\psfrag{x04}[t][t]{$10^{1}$}%
\psfrag{x05}[t][t]{$10^{-2}$}%
\psfrag{x06}[t][t]{$10^{-1}$}%
\psfrag{x07}[t][t]{$10^{0}$}%
\psfrag{x08}[t][t]{$10^{1}$}%
%
% yticklabels:
\psfrag{v01}[l][l]{0.016}%
\psfrag{v02}[l][l]{0.019}%
\psfrag{v03}[l][l]{0.022}%
\psfrag{v04}[l][l]{0.026}%
\psfrag{v05}[l][l]{0.029}%
\psfrag{v06}[l][l]{0.032}%
\psfrag{v07}[r][r]{2.58}%
\psfrag{v08}[r][r]{2.86}%
\psfrag{v09}[r][r]{3.15}%
\psfrag{v10}[r][r]{3.43}%
\psfrag{v11}[r][r]{3.71}%
%
% Figure:
%\resizebox{7cm}{!}{\includegraphics{fig_thr_est_time_tradeoff_fading.eps}}%
%\end{psfrags}%
%
% End fig_thr_est_time_tradeoff_fading.tex

\centering
\begin{tikzpicture}[scale=1]
\node[anchor=south west,inner sep=0] (image) at (0,0)
{
        %\includegraphics[trim=0.6cm 0.4cm 0.0cm 1.2cm,clip=true,width=\columnwidth]{figures/fig_thr_est_time_tradeoff_fading}
        %% eps trial
	\includegraphics[width= \columnwidth]{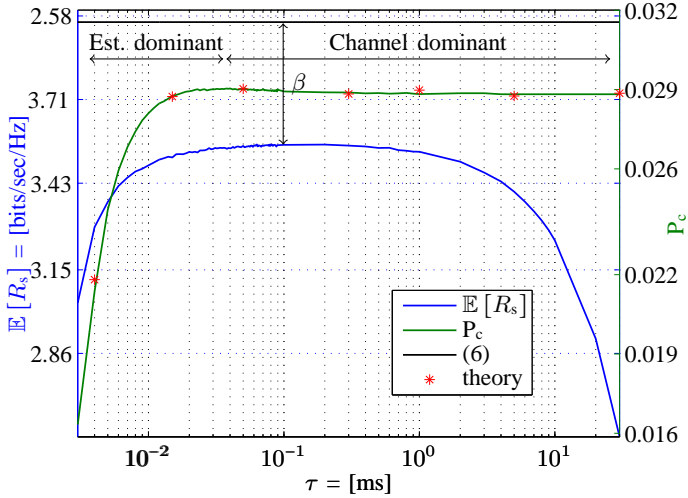}
};
\begin{scope}[x={(image.south east)},y={(image.north west)}]

% Select curves depending upon theta   
                        \draw[black,<->] (0.110,0.86) --  node[above = 0.0mm, font=\small] {Est. dominant} (0.31,0.86);
                        \draw[black,<->] (0.315,0.86) --  node[above = 0.0mm, font=\small] {Channel dominant} (0.89,0.86);

\draw[black,<->] (0.4,0.687) --  node[right=0.0mm, font=\small] {$\beta$} (0.4,0.932);

%\draw[help lines,xstep=.1,ystep=.1] (0,0) grid (1,1);
%\foreach \x in {0,1,...,9} { \node [anchor=north] at (\x/10,0) {0.\x}; }
%\foreach \y in {0,1,...,9} { \node [anchor=east] at (0,\y/10) {0.\y}; }
\end{scope}
\end{tikzpicture}
\caption{An illustration of estimation-throughput tradeoff for the channel with fading for $\gamma = \SI{0}{dB}$. $\beta$ represents the performance loss, however for this case the constraint for $\pc$ is not satisfied.}
\label{fig:ET_f}
\vspace{-5mm}
\end{figure}

%\begin{figure}[!t]
%        \includegraphics[trim=0.6cm 0.4cm 0.0cm 1.2cm,clip=true,width=\columnwidth]{figures/fig_thr_est_time_tradeoff_fading}
%\caption{An illustration of estimation-throughput tradeoff for the channel with fading for different $\gamma = [-10, 0]\SI{}{dB}$.}
%\label{fig:ET_f}
%\end{figure}
\figurename~\ref{fig:ET_f} captures the estimation throughput tradeoff for US with fading for $\gamma = \SI{0}{dB}$. It is interesting to note that for $\tau > \SI{30}{\us}$, $\pc$ attains saturation. Unlike path loss, despite increasing $\tau$, the variations in the $\pp$ across the $\ite$ at the PR cannot be reduced. This phenomenon is due to the inclusion of fading channel in power control, cf. (\ref{eq:preg}). Due to the influence of fading, we divide the system performance at the PR into estimation dominant regime and channel dominant regime, cf. \figurename~\ref{fig:ET_f}. In the estimation dominant regime $\tau < \SI{30}{\us}$, the system shows a improvement in $\pc$ with increase in $\tau$. In the channel dominant regime $\tau > \SI{30}{\us}$, $\pc$ reaches a saturation. This helps us to conclude that the variations in $\pp$ are mainly because of channel. Therefore, unlike the path loss channel, increasing $\tau$ doesn't improve the performance of the system. %However, the performance gets slightly better with $\gamma$, that is a higher $\gamma$ shifts the saturation point upwards. 

%Analogous to path loss channel, the performance at SR is depicted through $\eps$. 
Finally, we consider the performance at the SR in terms of expected throughput. For $\tau < \SI{20}{\us}$, the system experiences a strong change in $\ers$, it remains relatively constant for $\SI{20}{\us} < \tau < \SI{1}{ms}$ and finally decreases linearly beyond \SI{1}{ms}. To describe this behaviour, we consider (\ref{eq:Thr_pm}). Clearly, $\ers$ is dependent on $\gs$, $\preg$ and $(T- \tau)/T$, but only $\preg$ and $\tau$ depends on $N$. For $\tau < \SI{20}{\us}$, a slight change in $N$ has large influence on the value of $K$, cf. \figurename~\ref{fig:ET_f}. Thus, $K$ increases first drastically with $N$ and saturates after a certain point $\tau > \SI{20}{\us}$. However, beyond this point, the factor $(T- \tau)/T$ becomes dominant.

\section{Conclusion} \label{sec:conc}
%%%%%%%%%%%%%%%%%%%%%%%%%%%%%%%%%%%%%%%%%%%%%%%%%%%%%%%%%%%%%%%%%%%%%%%%%%%%%%%%%%%%%%%%%
%Power control is a prime requirement for the underlay cognitive radio. 
To implement power control mechanism for the US, it requires the knowledge of the channel at the ST. Acquiring the knowledge of channel through estimation induces variations in the parameters, e.g., power received at the PR. This may be harmful for the system. In this regard, we proposed a model to capture these variations and characterize the true performance of the US. The performance at PR and SR has been depicted based on probability of confidence and expected throughput. Moreover, the paper investigated the estimation-throughput tradeoff for the US. Based on analytical expressions, the maximum expected throughput for the secondary link was determined. 
Additionally, the variation in expected throughput against different system parameters was analyzed.
%More specifically, the performance bounds for the US were established. 
Finally, by means of numerical analysis, it has been clearly demonstrated that the conventional model overestimates the performance of the US.

For the underlay scenario with fading, it was observed that probability of confidence didn't improve despite increase in estimation time. Hence for fading channel, the probability of confidence is not a suitable choice for the PR constraint. For the future work, we would like to investigate estimation-throughput tradeoff for the US subject to the probability outage constraint at the PR. %Moreover, it is assumed in the model that the scaling factor and the path loss required for power regulation are perfectly known. However, in a real scenario these parameters must be estimated power several realizations. The estimation error due to that is not included in the system model. 

%%%%%%%%%%%%%%%%%%%%%%%%%%%%%%%%%%%%%%%%%%%%%%%%%%%%%%%%%%%%%%%%%%%%%%%%%%%%%%%%%%%%%%%%%
% References
%%%%%%%%%%%%%%%%%%%%%%%%%%%%%%%%%%%%%%%%%%%%%%%%%%%%%%%%%%%%%%%%%%%%%%%%%%%%%%%%%%%%%%%%%
\bibliographystyle{IEEEtran}
\bibliography{IEEEabrv,refs}

% that's all from my side
\end{document}